# The equivalent medium theory within the linear response region

Li-Gang Wang [1,2]

[1] *Department of Physics, Zhejiang University, Hangzhou, 310027, China*
[2] *Department of Physics, The Chinese University of Hong Kong, Shatin, Hong Kong*

In this paper, we present the equivalent medium theory by using the linear response theory. It is found that, under the condition of the linear response, a series of different media with different refractive indices $n_i(\omega)$ and lengths $d_i$ can be equivalent to an effective medium with the volume-averaged refractive index $\frac{1}{D}\sum_{i=1}^{N} n_i(\omega) d_i$ and the total length $D = \sum_{i=1}^{N} d_i$, where $N$ is the number of different media. Based on this equivalent theory, it is a simple but very useful method to design the effective medium with any desirable dispersion properties. As an example, we present a proposal to obtain the enhancement or reduction of the refractive index without absorption and the large dispersion without obvious absorption by assembling different linear dispersive gain and absorptive media.

OCIS index: 260. 2065, 260. 2030

It is well known that the dispersive properties [i. e., the index of refraction $n(\omega)$] of the media determine the propagation behaviors of a light pulse inside different dispersive media. Usually the dispersion and absorption (or gain) are fundamental characteristics of optical media. For many conventional media, the dispersion-absorption relation shows that at or near the resonance frequency, the index of refraction has a large dispersion but always companied with large absorption (or gain). Due to the large absorption or gain, the light pulse inside the medium is strongly distorted within a very short distance. Thus it has been an interesting subject for controlling the dispersion and absorption of the media [1-3].

The dispersion and absorption of the atomic systems can be dramatically manipulated by using quantum coherence and interference [2]. For example, Scully [14] in 1991 first proposed the idea of the enhancement of refractive index with vanishing absorption, by preparing the lower level doublet of the atom in a coherent superposition; Fleischhauer *et al.* [5, 6] studied various possible schemes to obtain the enhancement of the refractive index without absorption via atomic coherence and interference effects; Kocharovskaya *et al.* [7] suggested that a strong coherent field is used to drastically modify the spontaneous decay of three-level atoms, which leads to an anomalous atomic response yielding the modification of the dispersion and absorption of the atomic system. It is also well known that the dispersion of an atomic medium could be modified by the effect of electromagnetically induced transparency [8, 9], which has been applied to change the group velocity of a light pulse and realize the ultraslow propagation [10, 11]. Actually, in the last decades, much attention has been paid to the controlling of dispersion and absorption properties of an absorbing medium via various schemes using the effect of atomic coherence and interference with coherent or incoherent driving fields [12, 13]. However, all these schemes are usually involved in considerably complex experimental methods, which result in much expense or tremendous difficulties on the realization of the particular designed dispersive medium. For example, it is a complicated task to carry out the experiment about the enhancement of the refractive index by Zibrov *et al.* [14], where they achieved the change of the refractive index $\Delta n \approx 10^{-4}$. Until 2002,

Zibrov *et al.* [15] further observed the maximum resonant change of the refractive index about $\Delta n \approx 0.1$. Another example is that it was until 2001 that Wang et al. [16] used the active medium with double Raman gain peaks to realize the large anomalous dispersive region without large gain or absorption although the earlier idea was already presented by Bolda *et al.*[17] in 1994. In 2003, Stenner, Gauthier and Neifeld [18] ingeniously used two cells (of lengths $L/2$) containing the potassium vapors, which are respectively pumped by different laser beams with different frequencies, as an equivalent medium of length $L$ with a strong anomalous dispersion region. Inspired by the above factors, in the present paper, we present that, under the condition of the linear response, a combination of different media with different refractive indices $n_i(\omega)$ and lengths $d_i$ could be equivalent to an effective medium with the volume-averaged refractive index $\frac{1}{D}\sum_{i=1}^{N} n_i(\omega)d_i$ and the total length $D = \sum_{i=1}^{N} d_i$, where $N$ is the number of different media. This equivalent medium theory is effective within the linear response region. Using this equivalent theory, it is a simple but very powerful tool to design the effective medium with any desirable dispersion properties.

Consider the propagation of a light pulse passing through a linear dispersive medium with the refractive index $n(\omega) = \sqrt{\varepsilon(\omega)}$, where $\varepsilon(\omega)$ is the dielectric function. For the simplicity, we assume that the reflection from the linear dispersive medium is neglected, and the initial pulse $E(0,t)$ is incident from the position $z = 0$. According to the linear response theory, the pulse in the linear medium, at any position $z > 0$, can be given by [19]

$$E(z,t) = \frac{1}{2\pi} \int_{-\infty}^{\infty} E(0,\omega) \exp\left[i\frac{\omega n(\omega) z}{c} - i\omega t\right] d\omega, \quad (1)$$

where $E(0,\omega)$ is the initial spectrum of the incident pulse, and it could be obtained from the pulse intensity profile by using a Fourier integral: $E(0,\omega) = \int_{-\infty}^{\infty} E(0,t) \exp[i\omega t] d\omega$.

Now let us derive the equivalent medium theory of different linear dispersive media by using the linear response theory, as shown in Fig. 1, where $n_1(\omega)$, $n_2(\omega), \cdots, n_N(\omega)$ denote the refractive indices and $d_i$ denote lengths. From Eq. (1), the output pulse after the medium 1 could be easily written as

$$E_1(d_1,t) = \frac{1}{2\pi} \int_{-\infty}^{\infty} E(0,\omega) \exp\left[i\frac{\omega n_1(\omega)d_1}{c} - i\omega t\right] d\omega. \tag{2}$$

Obviously, the spectrum of the output pulse after the medium 1 (at position $z = d_1$) becomes

$$E_1(d_1,\omega) = E(0,\omega) \exp\left[i\frac{\omega n_1(\omega)d_1}{c}\right]. \tag{3}$$

Similarly the output pulse after the medium 2 [at position $z = (d_1 + d_2) + s_1$] could be readily obtained as follows,

$$E_2((d_1+d_2)+s_1,t) = \frac{1}{2\pi} \int_{-\infty}^{\infty} E(0,\omega) \exp\left[i\frac{\omega \frac{[n_1(\omega)d_1 + n_2(\omega)d_2]}{d_1+d_2} \times (d_1+d_2) + \omega s_1}{c} - i\omega t\right] d\omega, \tag{4}$$

and the corresponding spectrum could be expressed as

$$\begin{aligned}
E_2((d_1+d_2)+s_1,\omega) &= E_1(d_1,\omega) \exp\left[i\frac{\omega s_1 + \omega n_2(\omega)d_2}{c}\right], \\
&= E(0,\omega) \exp\left\{i\frac{\omega \frac{[n_1(\omega)d_1 + n_2(\omega)d_2]}{d_1+d_2} \times (d_1+d_2)}{c}\right\} \exp\left[i\frac{\omega s_1}{c}\right],
\end{aligned} \tag{5}$$

where $s_1$ is the vacuum distance between the medium 1 and the medium 2. Therefore, from Eq. (5), the medium 1 and the medium 2 can be equivalent to an effective medium with the volume averaged refractive index $\tilde{n}(\omega) = \frac{[n_1(\omega)d_1 + n_2(\omega)d_2]}{d_1+d_2}$ and total length $(d_1 + d_2)$.

By the similar steps, the output pulse after the medium $N$ as shown in Fig. 1 can be written as

$$\begin{aligned}
&E_N((d_1+d_2+\cdots+d_N)+(s_1+s_2+\cdots+s_{N-1}),t) \\
&= \frac{1}{2\pi} \int_{-\infty}^{\infty} E(0,\omega) \\
&\times \exp\left[i\frac{\omega \frac{[n_1(\omega)d_1 + n_2(\omega)d_2 + \cdots n_N(\omega)d_N]}{d_1+d_2+\cdots+d_N} \times (d_1+d_2+\cdots+d_N) + \omega(s_1+s_2+\cdots+s_{N-1})}{c} - i\omega t\right] d\omega.
\end{aligned} \tag{6}$$

From Eq. (6), it is found that *a series of linear dispersive media with different refractive indices $n_1(\omega)$, $n_2(\omega), \cdots, n_N(\omega)$ and lengths $d_i$ can be equivalent to an effective medium with the volume-averaged refractive index $\frac{1}{D}\sum_{i=1}^{N} n_i(\omega)d_i$ and the total length $D = \sum_{i=1}^{N} d_i$* [see Fig. 1(b)]. Using this equivalent medium theory, we can readily design any medium with any desirable dispersive properties by combining a series of simple linear dispersive media.

As an application example, we now apply this equivalent medium theory to obtain the enhancement (or reduction) of the refractive index and the large abnormal dispersive properties without obvious absorption. We consider a light pulse sequentially passing through five different linear media with different dielectric functions $\varepsilon_i(\omega)$ and the same lengths ($d_1 = d_2 = d_3 = d_4 = d_5 = d$). The dielectric functions are related with the linear susceptibilities $\chi(\omega)$ produced by two-level atoms, which may be easily controlled by different parameters. The dielectric functions for different media of two-level atoms under different parameters could be expressed as:

$$\begin{aligned}\varepsilon_i(\omega) &= 1 + \chi_i(\omega) \\ &= 1 + \frac{M_i}{\omega - \omega_0 - \Delta_i + i\gamma_i},\end{aligned} \quad (7)$$

where the subscripts $i = 1,2,3,4,5$ denote five different media, $\Delta_i = \omega_{0i} - \omega_0$ corresponds to the detuning of the resonant frequency $\omega_{0i}$ of the medium $i$ relative to the pulse's carrier frequency $\omega_0$, $M_i$ is proportional to the oscillator strength depending on both the transition dipole moment and the density of the two-level atoms: for $M_i > 0$, it means the gain two-level atoms, and for $M_i < 0$, it means the absorptive two-level atoms, and $\gamma_i$ the absorption or gain line-width. In this example, the equivalent medium has the volume-averaged refractive index $\frac{1}{5}\sum_{i=1}^{5} n_i(\omega)$ and the total length $5d$. In order to obtain the features of the enhancement (or reduction) of the refractive index and the large dispersion without obvious absorption, we take five different media under the following controlling parameters: $M_1 = M_3 = -1$ Hz, $M_2 = M_4 = 1$ Hz, $M_5 = -0.01272$ Hz, $\gamma_1 = \gamma_2 = \gamma_3 = \gamma_4 = 0.5$ MHz, $\gamma_5 = 0.1$ MHz, $\Delta_1 = 3$ MHz, $\Delta_2 = 2$ MHz, $\Delta_3 = -3$ MHz, $\Delta_4 = -2$ MHz, $\Delta_5 = 0$.

In Fig. 2 (a) we plot the real and imaginary parts of the refractive indices $n_i(\omega)$ ($i = 1,2,3,4,5$) for different media. It is clear seen that the media ①, ③ and ⑤ are absorptive anomalous dispersive media, and the media ② and ④ are gain normal dispersive media, and all media have different resonant frequencies. Obviously, such two-level atomic resonant dispersive media, also known as the Lorentz media, are readily controllably realized [17, 20, 21]. Figure 2(b) shows the equivalent dispersive curve of the equivalent medium for these five media. It is found that the equivalent refractive index could be enhanced greatly without obvious absorption [see the point A in Fig. 2(b)], and it is also possible to find that the averaged refractive index may be greatly reduced without obvious absorption [see the point C in Fig. 2(b)]. Finally it is also possible for obtaining the steep anomalous dispersive region without obvious absorption by suitably adjusting the parameters of the medium ⑤ [see the point B in Fig. 2(b)], and the detail anomalous dispersive curve near the point B is shown in Fig. 2(c).

It should be further pointed out that, in the above example, if we only combine the gain media ② and ④ without others, although these two media are normal dispersive, the combined effect of the media ② and ④ leads to a strong anomalous dispersive region with less gain. This situation is similar as in Ref. [18]. Apparently, using such linear medium's equivalent theory, it makes for easier designing the effective medium with the desirable dispersion properties.

In summary, we have derived the equivalent medium theory within the linear response region. It shows that a series of different media with refractive indices $n_i(\omega)$ and lengths $d_i$ could be effectively seen as a new medium with the volume-averaged refractive index $\frac{1}{D}\sum_{i=1}^{N} n_i(\omega)d_i$ and total length $D = \sum_{i=1}^{N} d_i$. In order to show the advantage of this equivalent medium theory, we present an example for obtaining the enhancement or reduction of the refractive index without absorption and the large dispersion without obvious absorption by assembling different linear dispersive gain and loss media. Using this equivalent theory, it is a simple but very effective method to design the effective medium with special-function dispersive properties. Finally, it should be pointed out that our study does not include the effect of the reflection from the interfaces between different media and the saturation effect of the media, which can be the researching subject of future work.

This work was supported by the National Nature Science Foundation of China (10604047 and 10547138), by Zhejiang Province Scientific Research Foundation (G20630 and G80611) and by the financial support from RGC of HK Government. L. G. Wang's email is lgwang@phy.cuhk.edu.hk.

# FIGURE CAPTIONS

FIG. 1. (a) A series of media with the different refractive indices $n_1(\omega)$, $n_2(\omega)$,... $n_N(\omega)$ and lengths $d_i$, separated by the vacuum with the distance $s_i$; (b) the equivalent medium with the volume-averaged refractive index $\frac{1}{D}\sum_{i=1}^{N} n_i(\omega)d_i$ and total length $D = \sum_{i=1}^{N} d_i$ and the vacuum with length $\sum_{i=1}^{N-1} s_i$.

FIG. 2. (Color Online) Real (solid line, $\text{Re}[n_i(\omega)]-1$) and imaginary (dashed line, $\text{Im}[n_i(\omega)]$) parts of (a) the refractive indices $n_i(\omega)$ of five different media and (b) the volume-averaged refractive index $\tilde{n}(\omega)$ of the equivalent medium. In (b), point A is the enhancement of the refractive index without absorption, point B the large anomalous dispersion without absorption, and point C the reduction of the refractive index without absorption. Plot (c) shows the real and imaginary parts of the averaged refractive index near the point B.

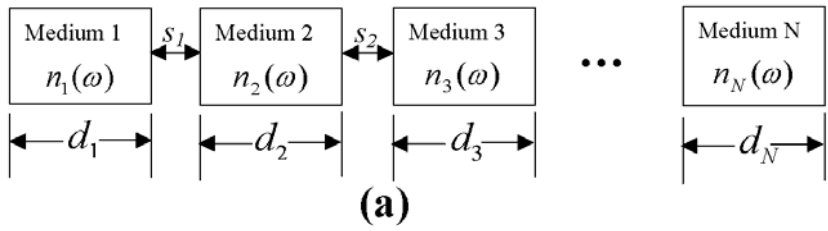

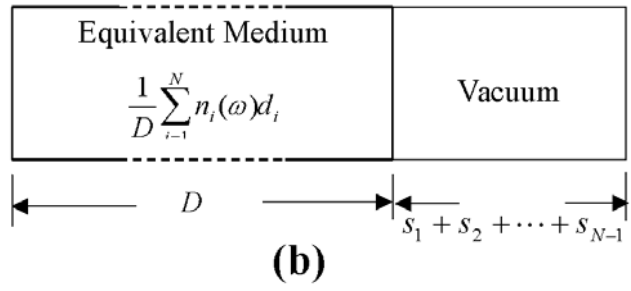

**FIG. 1**

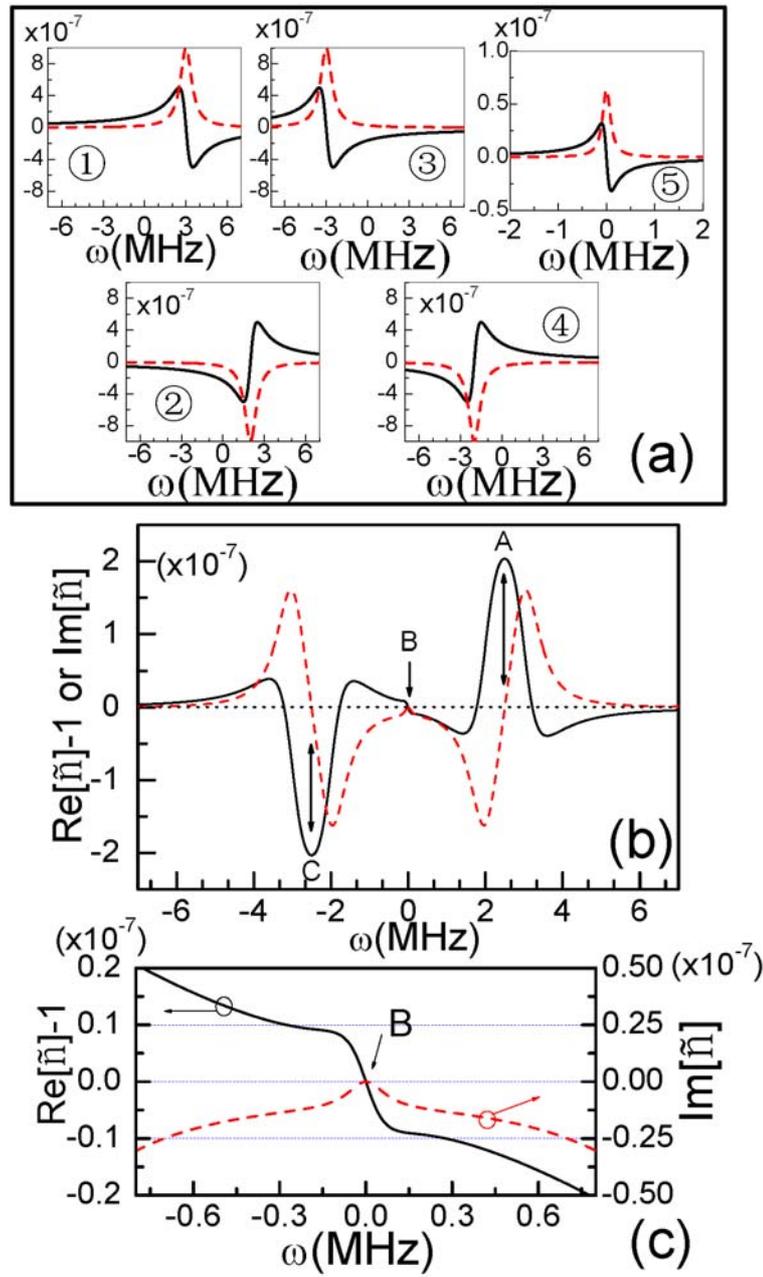

**Fig. 2**